\documentclass[11pt,twoside]{article}
\usepackage{amsmath}
\usepackage{amssymb}
\usepackage{amscd}
\usepackage{graphicx}
\usepackage{epstopdf}
\def\EndBox#1{
	\hskip0.1em\hfill\null\ \null\nobreak\hfill\kern3pt
		\hbox{$\scriptstyle #1$} \smallbreak}
\def\qed{\EndBox{\square}}

\newtheorem{proposition}{Proposition}[section]
\newcommand{\proof}{{\sc proof:~}}
\newcommand{\remark}{\smallbreak\noindent{\bf Remark.}}
\renewcommand{\a}{\alpha}
\renewcommand{\b}{\beta}
\newcommand{\g}{\gamma}
\newcommand{\G}{\Gamma}
\renewcommand{\d}{\delta}
\newcommand{\e}{\varepsilon}
\newcommand{\z}{{\zeta}}
\newcommand{\ze}{{\zeta}}
\renewcommand{\th}{\theta}
\newcommand{\Th}{\Theta}
\renewcommand{\l}{\lambda}
\newcommand{\m}{\mu}
\newcommand{\n}{\nu}
\renewcommand{\r}{\rho}
\newcommand{\s}{\sigma}
\renewcommand{\t}{\tau}

\newcommand{\om}{{\omega}}
\newcommand{\Cs}{{%
   \rlap{\lower3pt\hbox{\textnormal{\LARGE \char'040}}}{\Gamma}}{}}

\newcommand{\de}{\partial}
\newcommand{\oh}{\tfrac{1}{2}}
\newcommand{\ih}{\tfrac{\iO}{2}}
\newcommand{\oq}{\tfrac{1}{4}}
\newcommand{\osq}{\tfrac{1}{\surd2}}
\newcommand{\isq}{\tfrac{\iO}{\surd2}}
\newcommand{\cj}[1]{\overline{#1}}
\newcommand{\lin}{{\scriptscriptstyle\bigstar}}
\newcommand{\alin}{{\overline{\scriptscriptstyle\bigstar}}}
\renewcommand{\.}{{\scriptstyle\boldsymbol{\dot{}}}}
\newcommand{\td}{\tilde}
\newcommand{\up}{{\scriptscriptstyle\uparrow}}

\newcommand{\grav}{{}_{\mathrm{g}}}
\newcommand{\emag}{{}_{\mathrm{em}}}
\newcommand{\Dir}{{}_{\sst{\mathrm{D}}}}
\newcommand{\sbot}{{\scriptscriptstyle\bot}}
\newcommand{\ost}[1]{\overset{{}_{{\,}_*}}#1}
\newcommand{\fl}{\flat}
\newcommand{\bl}{{\bar\lambda}}
\newcommand{\bq}{{\bar q}}
\newcommand{\bs}{{\bar s}}
\newcommand{\bu}{{\bar u}}
\newcommand{\bv}{{\bar v}}
\newcommand{\bw}{{\bar w}}
\newcommand{\bzz}{{\bar\zz}}
\newcommand{\be}{{\bar\varepsilon}}
\newcommand{\bze}{{\bar\zeta}}
\newcommand{\bch}{{\bar\chi}}
\newcommand{\B}{{\boldsymbol{B}}}
\newcommand{\E}{{\boldsymbol{E}}}
\renewcommand{\H}{{\boldsymbol{H}}}
\renewcommand{\L}{{\boldsymbol{L}}}
\newcommand{\Ll}{{\scriptscriptstyle{\boldsymbol{L}}}}
\newcommand{\M}{{\boldsymbol{M}}}
\newcommand{\N}{{\boldsymbol{N}}}
\renewcommand{\P}{{\boldsymbol{P}}}
\newcommand{\Pm}{\P_{\!\!m}}
\newcommand{\Q}{{\boldsymbol{Q}}}
\newcommand{\Ql}{\Q{}^\lin}
\newcommand{\Qa}{\cj{\Q}{}^\lin}
\let\Sec=\S
\renewcommand{\S}{{\boldsymbol{S}}}
\newcommand{\Sc}{\cj{\S}}
\newcommand{\Sl}{\S{}^\lin}
\newcommand{\U}{{\boldsymbol{U}}}
\newcommand{\Uc}{\cj{\U}}

\newcommand{\Ua}{\cj{\U}{}^\lin}
\newcommand{\Ul}{\U{}^\lin}
\newcommand{\Uu}{{\scriptscriptstyle{\boldsymbol{U}}}}
\newcommand{\V}{{\boldsymbol{V}}}
\newcommand{\Vc}{\cj{\V}}

\newcommand{\Va}{\cj{\V}{}^\lin}
\newcommand{\Vl}{\V{}^\lin}
\newcommand{\W}{{\boldsymbol{W}}}
\newcommand{\Ww}{{\scriptscriptstyle{\boldsymbol{W}}}}
\newcommand{\Wc}{\cj{\W}}
\newcommand{\Wl}{\W{}^\lin}
\newcommand{\Lie}{\mathfrak{L}}
\newcommand{\Ug}{\mathrm{U}}

\newcommand{\SlG}{\mathrm{Sl}}

\newcommand{\Lor}{\mathrm{Lor}}

\newcommand{\Spin}{\mathrm{Spin}}
\newcommand{\CC}{{\mathbb{C}}}
\newcommand{\LL}{{\mathbb{L}}}
\newcommand{\QQ}{{\mathbb{Q}}}
\newcommand{\RR}{{\mathbb{R}}}
\newcommand{\UU}{{\mathbb{U}}}
\newcommand{\VV}{{\mathbb{V}}}
\newcommand{\ZZ}{{\mathbb{Z}}}
\newcommand{\Ccal}{{\mathcal{C}}}
\newcommand{\Lcal}{{\mathcal{L}}}
\newcommand{\Vcal}{{\mathcal{V}}}
\newcommand{\VC}{{\boldsymbol{\Vcal}}}
\newcommand{\End}{\operatorname{End}}
\newcommand{\Aut}{\operatorname{Aut}}
\newcommand{\Tr}{\operatorname{Tr}}
\newcommand{\Ker}{\operatorname{Ker}}
\newcommand{\Id}[1]{{1\!\!1}\!{}_{#1}{}}
\newcommand{\id}{{1\!\!1}}
\newcommand{\dO}{\mathrm{d}}
\newcommand{\DO}{\mathrm{D}}
\newcommand{\FO}{\mathrm{F}}
\newcommand{\kO}{\mathrm{k}}
\newcommand{\TO}{\mathrm{T}}
\newcommand{\TS}{\TO^{*}\!}
\newcommand{\dx}{\dO\xx}
\newcommand{\eO}{\mathrm{e}}
\newcommand{\iO}{\mathrm{i}}
\newcommand{\na}{\nabla\!}
\newcommand{\nasl}{{\rlap{\raise1pt\hbox{\,/}}\nabla}}
\newcommand{\ten}[1]{\operatorname*{\otimes}_{\!{\scriptscriptstyle #1}} }
\newcommand{\cart}[1]{\operatorname*{\times}_{\!{\scriptscriptstyle #1}} }
\newcommand{\dir}[1]{\operatorname*{\oplus}_{\!{\scriptscriptstyle #1}} }
\newcommand{\we}{{\,\wedge\,}}
\newcommand{\weu}[1]{{\wedge^{\!#1}}}
\newcommand{\pint}{\mathord{\rfloor}}
\newcommand{\comp}{\mathbin{\raisebox{1pt}{$\scriptstyle\circ$}}}
\newcommand{\tn}{{\,\otimes\,}}
\newcommand{\vh}{{\,\bar\vee\,}}
\newcommand{\bang}[1]{{\langle#1\rangle}}
\newcommand{\Ii}[2]{{}^{#1}_{\phantom{#1}\!#2}}
\newcommand{\iI}[2]{{}_{#1}^{\phantom{#1}\!#2}}
\newcommand{\iIi}[3]{{}_{#1\phantom{#2}\!\!#3}^{\phantom{#1}\!#2}}

\newcommand{\sA}{{\scriptscriptstyle A}}
\newcommand{\sB}{{\scriptscriptstyle B}}

\newcommand{\cA}{{\sA\.}}
\newcommand{\cB}{{\sB\.}}

\newcommand{\AAd}{{\sA\cA}}
\newcommand{\BBd}{{\sB\cB}}

\newcommand{\zeA}{{\zeta_\sA}}
\newcommand{\bzeA}{{\bze_\cA}}
\newcommand{\zeB}{{\zeta_\sB}}

\newcommand{\zzA}{\zz^\sA}
\newcommand{\zzB}{\zz^\sB}

\newcommand{\bzzB}{\bzz^\cB}

\newcommand{\Bsf}{{\mathsf{B}}}
\newcommand{\bb}{{\mathsf{b}}}
\renewcommand{\tt}{{\mathsf{t}}}
\newcommand{\uu}{{\mathsf{u}}}
\newcommand{\vv}{{\mathsf{v}}}
\newcommand{\ww}{{\mathsf{w}}}
\newcommand{\xx}{{\mathsf{x}}}
\newcommand{\zz}{{\mathsf{z}}}
\newcommand{\ie}{i.e$.$}
\newcommand{\eg}{e.g$.$}

\newcommand{\sst}{\scriptscriptstyle}
\newcommand{\sTh}{{\breve\Theta}}
\newcommand{\qRq}{{\quad\Rightarrow\quad}}
\newcommand{\gto}{\leadsto}

\newcommand{\onto}{\rightarrowtail}
\newcommand{\HL}{\H_{\!\Ll}}
\newcommand{\TOL}{\TO_{\!\Ll}}
\newcommand{\UL}{\U_{\!\!\Ll}}
\newcommand{\WL}{\W_{\!\!\!\Ll}}
\newcommand{\GF}{\G_{\!\!\sst\FO}}
\newcommand{\CF}{\Cs_{\!\!\sst\FO}}
\newcommand{\bCF}{\bar\Cs{}_{\!\!\sst\FO}}
\newcommand{\obCF}{\ost{{\bar\Cs}}{}_{\!\!\sst\FO}}
\title{Fermi transport of spinors and free QED states\\
in curved spacetime}
\date{{\small 25 November 2008} }
\author{Daniel Canarutto\\[6pt]
{\small\it Dipartimento di Matematica Applicata ``G. Sansone'', }\\
{\small\it Via S. Marta 3, 50139 Firenze, Italia}\\
{\small email:~daniel.canarutto@unifi.it}\\
{\small http://www.dma.unifi.it/\char126 canarutto}}
\begin{document}
\bibliographystyle{alpha}
\maketitle
\begin{abstract}
Fermi transport of spinors can be precisely understood
in terms of 2-spinor geometry.
By using a partly original, previously developed treatment
of 2-spinors and classical fields,
we describe the family of all transports, along a given
1-dimensional timelike submanifold of spacetime,
which yield the standard Fermi transport of vectors.
Moreover we show that this family has a distinguished member,
whose relation to the Fermi transport of vectors
is similar to the relation between the spinor connection
and spacetime connection.
Various properties of the Fermi transport of spinors
are discussed,
and applied to the construction of free electron states
for a detector-dependent QED formalism
introduced in a previous paper.
\end{abstract}

\noindent
2000 MSC:
53B05, 
53B21, 
81Q99. 

\noindent
Keywords:
Fermi transport, 2-spinors, Dirac spinors, free electron states.

\tableofcontents

\section{Two-spinors and Dirac spinors}\label{S:Two-spinors and Dirac spinors}

This section and the next one contain a sketch
of the two-spinor approach to Dirac algebra and field theories
referred to in the Introduction.
See~\cite{C98, C00b, C07} for details.

\subsection{Hermitian tensors}\label{s:Hermitian tensors}

If $\V$ is a finite dimensional complex vector space,
then we indicate by $\Vl$ its dual space,
by $\Va\cong\V^\alin$ its \emph{anti-dual space}
(namely the vector space of all anti-linear maps $\V\to\CC$)
and by $\Vc\cong\Vl{}^\alin\cong\Va{}^\lin$ its \emph{conjugate space}.
One then has natural anti-isomorphisms
$\V\to\Vc:v\mapsto\bv$ and $\Vl\to\Va:\l\mapsto\bl$\,.
Following a rather standard usage, we use ``dotted indices''
for vector and tensor components in $\Vc$ and $\Va$.

The space $\V\tn\Vc$ has a natural real linear (complex anti-linear) involution
$w\mapsto w^\dag$,
which on decomposable tensors reads
$$(u\tn\bv)^\dag=v\tn\bu~.$$
Hence one has the natural decomposition of $\V\tn\Vc$ into the direct sum
of the \emph{real} eigenspaces of the involution with eigenvalues $\pm1$,
respectively called the \emph{Hermitian} and \emph{anti-Hermitian} subspaces,
namely
$$\V\tn\Vc=(\V\vh\Vc)\oplus \iO\,(\V\vh\Vc)~.$$
In other terms, the Hermitian subspace $\V\vh\Vc$ is constituted by
all $w\in\V\tn\Vc$ such that $w^\dag=w$,
while an arbitrary $w$ is uniquely decomposed into the sum of an Hermitian
and an anti-Hermitian tensor as
$$w=\oh(w+w^\dag)+\oh(w-w^\dag)~.$$
In terms of components in any basis,
$w=w^{\sA\cB}\bb_\sA\tn\bar\bb_\cB$ is Hermitian (anti-Hermitian)
iff the matrix $(w^{\sA\cB}\,)$ of its components is of the same type,
namely $\bw^{\cB\sA}=\pm w^{\sA\cB}$.

\subsection{Two-spinor space}\label{s:Two-spinor space}

Let $\S$ be a $2$-dimensional complex vector space.
Then $\weu{2}\S$ is a $1$-dimensional complex vector space.
The Hermitian subspace of $(\weu{2}\S)\tn(\weu{2}\Sc)$
is a 1-dimensional real vector space with a distinguished orientation,
whose positively oriented semispace
$$\LL^2:=[(\weu{2}\S)\vh(\weu{2}\Sc)]^{+}:=\{w\tn\bw,~w\in\weu{2}\S\}$$
has the square root semi-space $\LL$,
called the space of \emph{length units}.\footnote{
A \emph{unit space} is defined to be a 1-dimensional real semi-space,
namely a positive semi-field associated with the semi-ring $\RR^+$
(see~\cite{JMV08} for details).
The \emph{square root} $\UU^{1/2}$ of a unit space $\UU$,
is defined by the condition that $\UU^{1/2}\tn\UU^{1/2}$ be isomorphic to $\UU$.
More generally, any \emph{rational power} of a unit space
is defined up to isomorphism
(negative powers correspond to dual spaces).
Here we only use the unit space $\LL$ of lengths and its powers;
essentially, this means that we take $\hbar=c=1$\,.}
The complex $2$-dimensional space
$$\U:=\LL^{-1/2}\tn\S$$
is called the \emph{$2$-spinor space}.
Observe that the $1$-dimensional space
$$\Q:=\weu{2}\U=\LL^{-1}\tn\weu{2}\S$$
has a distinguished Hermitian metric,
defined as the unity element in
$$\Qa\vh\Ql\equiv(\weu{2}\Ua)\vh(\weu{2}\Ul)
=\LL^{-2}\tn(\weu{2}\Sl)\vh(\weu{2}\Sl)\cong\RR~.$$
Hence there is the distinguished set of normalized ``symplectic'' forms on $\U$,
any two of them related by a phase factor.

Consider an arbitrary basis $(\xi_\sA)$ of $\S$\,,
and let $(\xx^\sA)$ be its dual basis of $\Sl$.
This determines the mutually dual bases
$$\ww:=\e^{\sA\sB}\,\xi_\sA\we\xi_\sB~,\quad
\ww^{-1}:=\e_{\sA\sB}\,\xx^\sA\we\xx^\sB~,$$
respectively of $\weu{2}\S$ and $\weu{2}\Sl$
(here $\e^{\sA\sB}$ and $\e_{\sA\sB}$ both denote
the antisymmetric Ricci matrix),
and the basis 
$$l:=\sqrt{\ww\tn\bar\ww} \quad\text{of}\quad \LL~.$$
Then one also has the induced mutually dual, {\em normalized\/} bases
$$(\zeA):=(l^{-1/2}\tn\xi_\sA)~,\quad (\zzA):=(l^{1/2}\tn\xx^\sA)$$
of $\U$ and $\Ul$, and also
\begin{align*}
&\e:=l\tn\ww^{-1}=\e_{\sA\sB}\,\zzA\we\zzB\in\Ql\equiv\weu{2}\Ul~, \\[6pt]
&\e^{-1}\equiv l^{-1}\tn\ww=\e^{\sA\sB}\,\zeA\we\zeB\in\Q\equiv\weu{2}\U~.
\end{align*}

\remark~
In contrast to the usual $2$-spinor formalism,
no symplectic form is fixed.
The  $2$-form $\e$ is unique up to a phase factor
which depends on the chosen 2-spinor basis,
and determines isomorphisms
\begin{align*}
& \e^\fl:\U\to\Ul:u\mapsto u^\fl~,~~
\bang{u^\fl,v}:=\e(u,v)\qRq (u^\fl)_\sB=\e_{\sA\sB}\,v^\sA~,\\[6pt]
& \e^\#:\Ul\to\U:\l\mapsto\l^\#~,~~
\bang{\m,\l^\#}:=\e^{-1}(\l,\m)\qRq (\l^\#)^\sB=\e^{\sA\sB}\,\l_\sA~.
\end{align*}
\smallbreak

\subsection{From 2-spinors to Minkowski space}
\label{s:From 2-spinors to Minkowski space}

Though a normalized element $\e\in\Ql$ is unique only up to a phase factor,
the tensor product $g\equiv\e\tn\be\in\Ql\tn\Qa$
is a naturally distinguished object.
This can also be seen as a bilinear form on $\U\tn\Uc$,
acting on decomposable elements as
$$g(p\tn\bq,r\tn\bs)=\e(p,r)\,\be(\bq,\bs)~.$$
The fact that any $\e$ is non-degenerate implies that $g$
is non-degenerate too.
In a normalized 2-spinor basis $(\zeA)$ one writes
$w=w^{\AAd}\,\zeA\tn\bzeA\in\U\tn\Uc$,
$g_{\AAd\,\BBd}=\e_{\sA\sB}\,\be_{\cA\cB}$ and
$$g(w,w)=\e_{\sA\sB}\,\be_{\cA\cB}\,w^{\AAd}\,w^{\BBd}=2\,\det w~.$$

The Hermitian subspace
$$\H:=\U\vh\Uc\subset\U\tn\Uc$$
is a $4$-dimensional \emph{real} vector space,
and the restriction of $g$ to $\H$
turns out to be a Lorentz metric with signature $(+,-,-,-)$\,.
Actually, for any given normalized basis $(\zeA)$ of $\U$ consider
the \emph{Pauli basis} $(\t_\l)$ of $\H$ associated with $(\zeA)$,
namely
$$\t_\l\equiv\t\iI{\l}{\AAd}\,\zeA\tn\bzeA
\equiv\osq\,\s\iI\l{\AAd}\,\zeA\tn\bzeA~,\quad \l=0,1,2,3~,$$
where $\s_\l$ denotes the $\l$-th Pauli matrix;
then one easily finds
$g(\t_\l\,,\t_\m)=2\,\d^0_\l\d^0_\m\,{-}\,\d_{\l\m}$\,.
Conversely, any orthonormal basis of $\H$ can be written as the Pauli basis
associated with an appropriate two-spinor basis.

It's not difficult to prove that
\emph{an element $w\in\U\tn\Uc=\CC\tn\H$ is null, that is $g(w,w)=0$\,,
iff it is a decomposable tensor: $w=u\tn\bs$, $u,s\in\U$\,.}
A null element in $\U\tn\Uc$ is also in $\H$
iff it is of the form $\pm u\tn\bu$.
Hence the \emph{null cone} $\N\subset\H$
is constituted exactly by such elements.
Note how this fact yields a way of distinguish between time orientations:
by convention, one chooses the \emph{future} and \emph{past}
null-cones in $\H$ to be, respectively,
$$\N^+:=\{u\tn\bu,~u\in\U\}~,\quad \N^-:=\{-u\tn\bu,~u\in\U\}~.$$

\subsection{From 2-spinors to Dirac spinors}
\label{s:From 2-spinors to Dirac spinors}

Next observe that an element of $\U\tn\Uc$
can be seen as a linear map $\Ua\to\U$,
while an element of $\Ua\tn\Ul$ can be seen as a linear map $\U\to\Ua$.
Then one defines the linear map
\begin{align*}
&\g:\U\tn\Uc\to\End(\U\oplus\Ua):y\mapsto
\g(y):=\sqrt2\,\bigl(y,y^{\fl\lin}\bigr)~,\phantom{\text{\ie}\quad}\\[6pt]
\text{\ie}\quad
&\g(y)(u,\chi)=\sqrt2\bigl(y\pint\chi\,,u\pint y^\fl \bigr)~,
\end{align*}
where $y^\fl:=g^\fl(y)\in\Ul\tn\Ua$ and $y^{\fl\lin}\in\Ua\tn\Ul$
is the transposed tensor.
In particular for a decomposable $y=p\tn\bq$ one has
$$\td\g(p\tn\bq)(u,\chi)
=\sqrt2\bigl(\bang{\chi,\bq}\,p\,,\bang{p^\fl,u}\,\bq^\fl\,\bigr)~.$$

It's not difficult to see that, for all $y,y'\in\U\tn\Uc$\,, one has
$$\g(y)\comp\g(y')+\g(y')\comp\g(y)=2\,g(y,y')\,\id~,$$
namely $\g$ is a \emph{Clifford map} relatively to $g$;
thus one is led to regard $$\W:=\U\oplus\Ua$$ as the space of Dirac spinors,
decomposed into its Weyl subspaces.
The restriction of $\g$ to the Minkowski space $\H$
is called the \emph{Dirac map}.

The 4-dimensional complex vector space $\W$ is naturally endowed
with a further structure:
the obvious anti-isomorphism
$$\W\to\Wl:(u,\chi)\mapsto(\bch,\bu)~.$$
Namely, if $\psi=(u,\chi)\in\W$ then $\bar\psi=(\bu,\bch)\in\Wc$
can be identified with $(\bch,\bu)\in\Wl$\,;
this is the so-called `Dirac adjoint' of $\psi$\,.
This operation can be seen as the ``index lowering anti-isomorphism''
related to the Hermitian product
$$\kO:\W\times\W\to\CC:\Bigl((u,\chi),(u',\chi')\Bigr)
\mapsto\bang{\bch,u'}+\bang{\chi',\bu}~,$$
which is obviously non-degenerate;
its signature turns out to be $(+\,+\,-\,-)$,
as it can be seen in a ``Dirac basis'' (below).

Let $(\zeA)$ be a normalized basis of $\U$\,;
the \emph{Weyl basis} of $\W$
is defined to be the basis $(\z_\a)$, $\a=1,2,3,4$, given by
$$(\z_1\,,\z_2\,,\z_3,\z_4):=(\z_1\,,\z_2\,,-\bzz^1,-\bzz^2)~.$$
The \emph{Dirac basis} $(\z'_\a)$, $\a=1,2,3,4$, is given by
\begin{align*}
& \z'_1=\osq(\z_1\,,\bzz^1)\equiv\osq(\z_1-\z_3)~,
&& \z'_2=\osq(\z_2\,,\bzz^2)\equiv(\z_2-\z_4)~,\\[6pt]
& \z'_3=\osq(\z_1\,,-\bzz^1)\equiv(\z_1+\z_3)~,
&& \z'_4=\osq(\z_2\,,-\bzz^2)\equiv(\z_2+\z_4)~.
\end{align*}
Setting
$$\g_\l:=\g(\t_\l)\in\End(\W)$$
one recovers the usual Weyl and Dirac representations as the matrices
$\bigl(\g_\l\bigr)$\,, $\l=0,1,2,3$\,,
in the Weyl and Dirac bases respectively.

It should be noted that no distinguished Hermitian metric exists
either on $\U$ or $\W$\,:
assigning such structure is equivalent to fixing an observer 
(this point remains somewhat obscured
in most traditional treatments of spinors).
In fact, a Hermitian 2-form $h$ on $\U$ is an element in $\Ua\vh\Ul\cong\H^*$.
One says that $h$ is \emph{normalized} if it is non-degenerate, positive
and $g^\#(h)=h^{-1}$;
the latter condition is equivalent to $g(h,h)=2$\,.
If $h$ is normalized then it is necessarily
a future-pointing timelike element in $\H^*$\,.
For example, if $(\t_\l)$ is a Pauli basis and $(\tt^\l)$ is the dual basis,
then $\sqrt2\,\bar\tt^0=\bzz^1\tn\zz^1+\bzz^2\tn\zz^2$ is normalized;
conversely, every positive-definite normalized Hermitian metric $h$
can be expressed in the above form for some suitable
normalized 2-spinor bases.
In 4-spinor terms:
if $h$ is assigned, then it extends naturally
to a Hermitian metric $h$ on $\W$,
which can be characterized by\footnote{
In the traditional notation,
$\g_\l^\dag$ indicates the $h$-adjoint of $\g_\l$\,,
and then depends on the chosen observer.} 
$$h(\psi,\phi)=\kO(\g_0\psi,\phi)~.$$

\remark~
Some other operations on 4-spinor space, commonly used in the literature,
actually depend on particular choices or conventions.
\emph{Charge conjugation}, in particular, is the antilinear involution
$$\Ccal:\W\to\W:(u,\chi)\mapsto
\eO^{-\iO t}\,\bigl(\e^\#(\bch),-\be^\fl(\bu)\bigr)$$
determined by the choice of a normalized delement
$\om\equiv\eO^{\iO\,t}\,\e\in\weu2\Ul$.
\emph{Parity} is the endomorphism $\g_0\equiv\g(\t_0)$\,,
so it depends on the choice of an observer
(here written as the element $\t_0$ of a suitable Pauli frame).
\emph{Time-reversal} is the composition $\gamma_\eta\,\g_0\,\Ccal$\,,
where $\gamma_\eta$ (in a Pauli basis: $\gamma_\eta=\g_0\,\g_1\,\g_2\,\g_3$)
is the endomorphism corresponding, via $\g$\,,
to the volume form $\eta$ determined by $g$ on $\H$.

\section{Two-spinor bundle and field theories}
\label{S:Two-spinor bundle and field theories}
\subsection{Two-spinor connections}\label{s:Two-spinor connections}

Consider any real manifold $\M$ and a vector bundle
$\S\onto\M$ with complex $2$-dimensional fibres.
Denote base manifold coordinates as $(\xx^a)$;
choose a local frame $(\xi_\sA)$ of $\S$,
determining linear fibre coordinates $(\xx^\sA)$.
According to the constructions of the previous sections,
one now has the bundles $\Q$, $\LL$, $\U$, $\H$, $\W$ over $\M$,
with smooth natural structures;
the frame $(\xi_\sA)$ yields the frames $\e$, $l$, $(\zeA)$ and $(\t_\l)$\,,
respectively.
Moreover for any rational number $r\in\QQ$
one has the semi-vector bundle $\LL^r$\,.

Consider an arbitrary $\CC$-linear connection $\Cs$ of $\S\onto\M$,
called a \emph{$2$-spinor connection}.
In the fibred coordinates $(\xx^a,\xx^\sA)$\, $\Cs$ is expressed
by the coefficients $\Cs\iIi{a}{\sA}{\sB}:\M\to\CC$\,,
namely the covariant derivative of a section $s:\M\to\S$ is expressed as
$$\nabla s=(\de_a s^\sA-\Cs\iIi{a}{\sA}{\sB}s^\sB)\,\dx^a\tn\xi_\sA~.$$
The rule
$\nabla\bs=\overline{\nabla s}$
yields a connection $\bar\Cs$ on $\Sc\onto\M$,
whose coefficients are given by
$$\bar\Cs\iIi{a}{\cA}{\cB}=\overline{\Cs\iIi{a}{\sA}{\sB}}~.$$
Actually, $\Cs$ determines linear connections on each of the above said
induced vector bundles over $\M$.
Denote by $2\,G$ and $2\,Y$ the connections induced on $\LL$ and $\Q$
(this notation makes sense because the fibres are 1-dimensional), namely
\begin{gather*}
\nabla l=-2\,G_a\,\dx^a\tn l~,\quad  \nabla\e=2\,\iO\,Y_a\,\dx^a\tn\e~,\\
\nabla\ww^{-1}\equiv\nabla(l^{-1}\tn\e)=2(G_a+\iO\,Y_a)\,\dx^a\tn l^{-1}\tn\e
\end{gather*}
and the like. By direct calculation we find
$$
G_a=\oq(\Cs\iIi{a}{\sA}{\sA}+\bar\Cs\iIi{a}{\cA}{\cA})~,\qquad
Y_a=\tfrac{1}{4\iO}(\Cs\iIi{a}{\sA}{\sA}-\bar\Cs\iIi{a}{\cA}{\cA})~.
$$
Since $Y_a$ is real, the induced linear connection on $\Q$
is Hermitian (preserves its natural Hermitian structure).

The coefficients of the induced connections $\td\Cs$ on $\U$,
and $\td\G$ on $\H$, turn out to be
\begin{align*}
&\td\Cs\iIi a\sA\sB=\Cs\iIi a\sA\sB-G_a\,\d\Ii\sA\sB~,
\\[6pt]
&\td\G\iIi{a}{\AAd}{\BBd}=
\Cs\iIi a\sA\sB\,\d\Ii\cA\cB+\d\Ii\sA\sB\,\bar\Cs\iIi a\cA\cB
	-2\,G_a\,\d\Ii\sA\sB\,\d\Ii\cA\cB~.
\end{align*}
Since its coefficients are real,
$\td\G$ turn out to be reducible to a real connection on $\H$.
Moreover this connection $\td\G$ turns out to be \emph{metric},
namely $\nabla[\td\G]g=0$\,.
Hence, its coefficients are antisymmetric and traceless, namely
$$\td\G\iI{a}{\l\m}+\td\G\iI{a}{\m\l}=0~,\quad\td\G\iIi{a}{\l}{\l}=0~.$$

The above relations between $\Cs$ and the induced connections
can be inverted as
$$\Cs\iIi a\sA\sB
=(G_a+\iO\,Y_a)\,\d\Ii\sA\sB+\oh\,\td\G\iIi a{\AAd}{\sB\cA}~,$$
and a similar relation holds among the curvature tensors, namely
$$R\iIi{ab}{\sA}{\sB}=
-2\,(\dO G+\iO\,\dO Y)_{ab}\,\d\Ii{\sA}{\sB}
+\oh\,\td R\iIi{ab}{\AAd}{\sB\cA}~.$$

\subsection{Two-spinor tetrad}\label{s:Two-spinor tetrad}

Henceforth I'll assume that $\M$ is a real $4$-dimensional manifold.
Consider a linear morphism
$$\Th:\TO\M\to\S\tn\Sc=\CC\tn\LL\tn\H~,$$
namely a section
$$\Th:\M\to\CC\tn\LL\tn\H\tn\TS\M$$
(all tensor products are over $\M$).
Its coordinate expression is
$$ \Th=\Th_a^\l\,\t_\l\tn\dx^a=\Th_a^\AAd\,\zeA\tn\bzeA\tn\dx^a~,
\qquad \Th_a^\l,\Th_a^\AAd:\M\to\CC\tn\LL~.$$

We'll assume that $\Th$ is non-degenerate
and valued in the Hermitian subspace $\LL\tn\H\subset\S\tn\Sc$\,;
then $\Th$ can be viewed as a `scaled' \emph{tetrad}
(or \emph{soldering form}, or \emph{vierbein});
the coefficients $\Th_a^\l$ are real (\ie\ valued in $\RR\tn\LL$)
while the coefficients $\Th_a^\AAd$ are Hermitian,
\ie\ $\bar\Th_a^{\cA\sA}=\Th_a^\AAd$.
Through a tetrad, the geometric structure of the fibres of $\H$
is carried to a similar, scaled structure on the fibres of $\TO\M$.
It will then be convenient, from now on, to distinguish by a tilda
the objects defined on $\H$,
so I'll denote by $\td g$\,, $\td\eta$ and $\td\g$
the Lorentz metric,
the $\td g$-normalized volume form and the Dirac map of $\H$\,, and set
\begin{align*}
g&:=\Th^*\td g:\M\to\CC\tn\LL^2\tn\TS\M\tn\TS\M~,
\\[6pt]
\eta&:=\Th^*\td\eta:\M\to\CC\tn\LL^4\tn\weu{4}\TS\M~,
\\[6pt]
\g&:=\td\g\comp\Th:\TO\M\to\LL\tn\End(\W)~,
\end{align*}
which have the coordinate expressions
\begin{align*}
g&=\eta_{\l\m}\,\Th_a^\l\,\Th_b^\m\,\dx^a\tn\dx^b
=\e_{\sA\sB}\e_{\cA\cB}\,\Th_a^\AAd\,\Th_b^\BBd\,\dx^a\tn\dx^b~,
\\[6pt]
\eta&=\det(\Th)\,\dx^0\we\dx^1\we\dx^2\we\dx^3~,
\\[6pt]
\g&=\sqrt2\,\Th_a^\AAd\,
(\zeA\tn\bzeA+\e_{\sA\sB}\e_{\cA\cB}\,\bzzB\tn\zzB)\tn\dx^a~.
\end{align*}
The above objects turn out to be a Lorentz metric,
the corresponding volume form and a Clifford map.
Moreover
$$ \Th_\m^b:=\Th_a^\l\,\eta_{\l\m}\,g^{ab}=(\Th^{-1})_\m^b:\M\to\CC\tn\LL^{-1}~,
\quad
g^{ab}:\M\to\CC\tn\LL^{-2}~.$$

A non-degenerate tetrad, together with a two-spinor frame,
yields mutually dual orthonormal frames
$(\Th_\l)$ of $\LL^{-1}\tn\TO\M$
and $(\ost\Th{}^\l)$ of $\LL\tn\TS\M$\,, given by
$$\Th_\l:=\Th^{-1}(\t_\l)=\Th_\l^a\,\de\xx_a~,\quad
\ost\Th{}^\l:=\Th^*(\tt^\l)=\Th_a^\l\,\dx^a~.$$

We also write
\begin{align*}&
\g=\g_\l\tn\ost\Th{}^\l=\g_a\tn\dx^a~,\quad
\g_\l:=\g(\Th_\l):\M\to\End(\W)~,\\&
\g_a:=\g(\de\xx_a)=\Th_a^\l\,\g_\l:\M\to\LL\tn\End(\W)~.
\end{align*}

If $\Cs$ is a complex-linear connection on $\S$,
and $G$ and $\td\G$ are the induced connections on $\LL$ and $\H$,
then a non-degenerate tetrad $\Th:\TO\M\to\LL\tn\H$ yields a unique
connection $\G$ on $\TO\M$, characterized by the condition
$$\nabla[\G\tn\td\G]\Th=0~.$$
Moreover $\G$ is metric, \ie\ $\nabla[\G]g=0$.
Denoting by $\G\iIi{a}{\l}{\m}$ the coefficients of $\G$
in the frame $\Th_\l'\equiv\Th^{-1}(l\tn\t_\l)$ one obtains
$$\G\iIi{a}{\l}{\m}=\td\G\iIi{a}{\l}{\m}+2\,G_a\,\d\Ii{\l}{\m}~.$$

The curvature tensors of $\G$ and $\td\G$ are related by
$R\iIi{ab}{\l}{\m}=\td R\iIi{ab}{\l}{\m}$\,, or
$$R\iIi{ab}{c}{d}=\td R\iIi{ab}{\l}{\m}\,\Th_\l^c\,\Th_d^\m~.$$
Hence the Ricci tensor and the scalar curvature are given by
$$R_{ad}=R\iIi{ab}{b}{d}=\td R\iIi{ab}{\l}{\m}\,\Th_\l^b\,\Th_d^\m~,\qquad
R\iI{a}{a}=\td R\iI{ab}{\l\m}\,\Th_\l^b\,\Th_\m^a~.$$

In general,
the connection $\G$ will have non-vanishing torsion,
which can be expressed as
$$\Th_c^\l\,T\Ii{c}{ab}=\de_{[a}^{\phantom{a}}\Th_{b]}^\l
+\Th_{[a}^\m\,\td\G\iIi{b]}{\l}{\m}
+2\,\Th_{[a}^\l\,G^{\phantom{a}}_{b]}~.$$

\subsection{Einstein-Cartan-Maxwell-Dirac field theory}
\label{s:Einstein-Cartan-Maxwell-Dirac field theory}

In this section I'll give an essential sketch
of a ``minimal geometric data'' field theory
which has been presented in previous papers~\cite{C98, C00b, C07}.
The quoted words refer to the fact that the unique ``geometric datum''
is a vector bundle $\S\onto\M$
with complex 2-dimensional fibres and real 4-dimensional base manifold.
All other bundles and fixed geometric objects are determined just
by this datum through functorial constructions,
as we saw in the previous sections;
no further background structure is assumed.
Any considered bundle section which is not functorially fixed
by our geometric datum is a field.
A natural Lagrangian can then be written, yielding a field theory
which turns out to be essentially equivalent to a classical theory
of Einstein-Cartan-Maxwell-Dirac fields.

The fields are taken to be the tetrad $\Th$\,,
the $2$-spinor connection $\Cs$,
the electromagnetic field $F$ and the electron field $\psi$\,.
The gravitational field is represented by $\Th$
(which can be viewed as a `square root' of the metric)
and the traceless part of $\Cs$, namely $\td\G$,
seen as the gravitational part of the connection.
If $\Th$ is non-degenerate one obtains,
as in the standard metric-affine approach~\cite{HCMN,FK82},
essentially the Einstein equation and the equation for torsion;
the metricity of the spacetime connection is a further consequence.
But note that the theory is non-singular also in the degenerate case.
The connection $G$ induced on $\LL$ will be assumed to have vanishing curvature,
$\dO G=0$, so that one can always find local charts such that $G_a=0$\,;
this amounts to gauging away the conformal (`dilaton') symmetry.
Coupling constants will arise as covariantly constant sections of $\LL$,
which now becomes just a vector space.

The Dirac field is a section
$\psi\equiv(u,\chi):\M\to\LL^{-3/2}\tn\W$
assumed to represent a semiclassical particle with one-half spin,
mass $m\in\LL^{-1}$ and charge $q\in\RR$\,.

The electromagnetic potential can be thought of as
the Hermitian connection on $\weu2\U$ determined by $\Cs$\,,
whose coefficients are indicated as $\iO\,Y_a$\,;
locally one writes
$Y_a\equiv q\,A_a$\,,
where $A:\M\to\TS\M$ is a local 1-form.

The electromagnetic field is represented by a spinor field
$\td F:\M\to\LL^{-2}\tn\weu2\H^*$
which, via $\Th$\,, determines the 2-form $F:=\Th^*\td F:\M\to\weu2\TS\M$\,.
The relation between $Y$ and $F$ will follow as one of the field equations.

The total Lagrangian density is the sum of
a gravitational, an electromagnetic and a Dirac term:
$\Lcal=\Lcal\grav+\Lcal\emag+\Lcal\Dir
=(\ell\grav+ \ell\emag+ \ell\Dir)\,\dO^4\xx:\M\to\weu4\TO^*\M$\,,
where
\begin{align*}
\Lcal\grav&:=
\frac{1}{8\,\Bbbk}\,\td\eta\mid(\td R^\#\we\Th\we\Th)~,
\\[6pt]
\Lcal\emag&:=
-\oh\,\td\eta\mid[\Th\we\Th\we(\dO A\tn\td F)]
+\oq\,(\td F{\cdot}\td F)\,\eta~,
\\[6pt]
\Lcal\Dir&:=\Im\bigl[\tfrac1{3!}\,
\bigl\langle\bar\psi\,,\;
\td\eta\mid(\td\g^\#\nabla\psi)\we\Th\we\Th\we\Th\bigr\rangle]
-m\,\bang{\bar\psi,\psi}\,\eta~.
\end{align*}
In the above expressions, $\Bbbk$ is Newton's gravitational constant;
a superscript $\#$ denotes ``index raising''
relatively to the Lorentz metric $\td g$\,,
and the ``exterior'' product among vector valued forms
is naturally defined.\footnote{
For example, $(x\tn\a)\we(y\tn\b)=(x\we y)\tn(\a\we\b)$\,,
$\a,\b\in\TS\M$, $x,y\in\H$\,, and the like.}
One has the coordinate expressions
\begin{align*}
\ell\grav&=\frac{1}{8\,\Bbbk}\,
\e_{\l\m\n\r}\,\e^{abcd}\,\td R\iI{ab}{\l\m}\,\Th_c^\n\,\Th_d^\r~,
\\[6pt]
\ell\emag&=-\oq\,\e^{abcd}\,\e_{\l\m\n\r}\,\de_aA_b\,\td F^{\l\m}\,\Th_c^\n\Th_d^\r
+\oq\,\td F^{\a\b}\td F_{\a\b}\,\det\Th~,
\\[6pt]
\ell\Dir&=\isq\,\sTh^a_{\AAd}\,\Bigl(\na_au^\sA\,\bu^\cA-u^\sA\,\na_a\bu^\cA
+\e^{\sA\sB}\be^{\cA\cB}(\bch_\sB\,\na_a\chi_\cB-\na_a\bch_\sB\,\chi_\cB\,)
\Bigr)
\\
&\hspace{6cm}-m\,(\bch_\sA u^\sA+\chi_\cA\,\bu^\cA\,)\,\det\Th~,
\end{align*}
where
$$\sTh^a_{\AAd}\equiv\osq\,\s\Ii\l{\AAd}\,\sTh^a_\l\equiv
\osq\,\s\Ii\l{\AAd}\,\bigl(
\tfrac{1}{3!}\,\e^{abcd}\,\e_{\l\m\n\r}\,\Th_b^\m\Th_c^\n\Th_d^\r\bigr)~.$$

Writing down the Euler-Lagrange equations\footnote{
One has to calculate the variational derivatives relatively to all the fields
$\Cs$, $\Th$\,, $A$\,, $\td F$, $\psi\equiv(u,\chi)$\,.}
for the Lagrangian density $\Lcal$
is a straightforward (though not short) task.
Summarizing the basic results:
\begin{list}{$\bullet$}{}
\item
The $\Th$-component corresponds
(in the non-degenerate case) to the Einstein equation.
\item
The $\td\G$-component gives the equation for torsion.
From this one sees that the spinor field is a source for torsion,
and that in this context one cannot formulate a torsion-free theory.
\item
The $\td F$-component reads $F=2\,\dO A$ in the non-degenerate case,
and of course this yields the first Maxwell equation $\dO F=0$.
\item
The $A$-component reduces, in the non-degenerate case,
to the second Maxwell equation
$\oh\,{*}\dO{*}F=j$\,,
where $j:\M\to\TS\M$ is the \emph{Dirac current}.
\item
The $\bu$- and $\bch$-components
give a generalized form of the standard \emph{Dirac equation},
which can be written in compact form as
$$\bigl(\iO\,\nasl-m+\ih\,\g^\#(\breve T)\bigr)\psi=0~.$$
Here, $\breve T$ denotes the $1$-form obtained from the torsion by contraction,
with coordinate expression $\breve T_a=T\Ii{b}{ab}$\,.
\end{list}

\section{Fermi transport}\label{S:Fermi transport}

A 1-dimensional timelike submanifold $\L\subset\M$
can be seen as a `pointlike observer',
or as the world-line of a `detector'.
Note that there is a natural inclusion $\TO\L\subset\TO\M$.
The restriction of the spacetime time metric is a Riemann metric on $\L$,
which yields the detecor's `proper time'.

Throughout this \Sec\ref{S:Fermi transport} we'll assume
a tetrad $\Th$ and a spinor connection $\Cs$ to be fixed,
namely we'll work in a given gravitational field background.
Moreover, for simplicity, we'll assume $G_a=0$ as it is
in the standard fied theories
(\Sec\ref{s:Einstein-Cartan-Maxwell-Dirac field theory}).

Since $\Th$ is fixed, it will be convenient
to make the identification $\TO\M\cong\LL\tn\H$ (and the like)
in order to simplify our notations.
Note (\Sec\ref{s:Two-spinor tetrad})
that the scalar product of elements in $\TO\M$
is valued into $\RR\tn\LL^2$,
while the tensor product of elements in $\H$ is real valued;
we express this fact by saying that the spacetime metric $g$
is \hbox{$\LL^2$-\emph{scaled}},
while the metric on $\H$ is \emph{unscaled}
(or `conformally invariant').

\subsection{Rivisitation of the standard Fermi transport}
\label{s:Rivisitation of the standard Fermi transport}

Denote as $\TOL\M\onto\L$ and $\HL\onto\L$ 
the restrictions of the bundles $\TO\M\onto\M$ and $\H\onto\M$
to the base $\L$
(the fibres over elements in $\L$ are the same).
Then $\HL\cong\LL^{-1}\tn\TOL\M$ has one distinguished section,
namely the unit future-pointing scaled vector field
$$\t:\L\to\LL^{-1}\tn\TO\L\subset\LL^{-1}\tn\TOL\M\cong\HL~.$$

We now consider the linear morphism over $\L$
$$\Phi:\TO\L\to\weu2\HL:v\mapsto\Phi_v\equiv 2\,(\na_v\t)\we\t~.$$
Choose base coordinates $(\xx^a)\equiv(\xx^1,\xx^2,\xx^3,\xx^4)$
\emph{adapted} to $\L$,
namely such that $\de\xx_4\equiv\de/\de\xx^4$
is tangent to $\L$ at the points of $\L$;
let moreover $(\t_\l)$ be any orthonormal frame of $\HL$
such that $\t_0\equiv\t$\,;
then one gets the coordinate expression
$$\Phi=2\,\td\G\iI 4{0j}\,\dx^4\tn\t_j\we\t_0~.$$

By `lowering the second index' of $\Phi$ through the metric
one gets a linear morphism 
$$\Phi^\fl:\L\to\HL\tn\HL^*\equiv\End(\HL)~,$$
namely
$$\Phi^\fl_v=\na_v\t_0\tn\tt^0-\t_0\tn\na_v\tt^0~,$$
with coordinate expression
$$\Phi^\fl=-\dx^4\tn(\td\G\iIi4j0\,\t_j\tn\tt^0+\td\G\iIi40j\,\t_0\tn\tt^j)~,$$
where the dual frame of $(\t_\l)$ was denoted as $(\tt^\l)$.

Note that
$$(\Phi^\fl)\iIi400=(\Phi^\fl)\iIi4jj=(\Phi^\fl)\iIi4\l\l=0~.$$

The bundle $\HL\onto\L$
has of course the connection naturally induced by $\td\G$\,:
the covariant derivative of any section $X:\L\to\HL$
is defined to be the map $v\mapsto\na_vX:\L\to\HL$\,,
namely $\na_vX\equiv\na_v[\td\G]X$ is the restriction of $\na_{v'}[\td\G]X'$
for \emph{any} local extensions $v'$ and $X'$ of $v$ and $X$\,.
But now we observe that $\Phi^\fl$ can be viewed
as a section $\L\to\TS\L\ten{\L}\End(\HL)$\,,
according to $\Phi^\fl_v\equiv v\pint\Phi^\fl$\,.
Hence we are able to introduce a new connection of $\HL\onto\L$\,,
namely the \emph{Fermi connection}\footnote{
We recall that the difference between any two connections
on a vector bundle $\E\onto\B$
is a tensor field \hbox{$\B\onto\TS\B\ten{\B}\End(\E)$}\,.}
$$\GF:=\td\G+\Phi^\fl~.$$
The covariant derivative associated with $\GF$
turns out to have the expression
\begin{align*}
\DO_vX &\equiv\na_v[\GF]X
=\na_vX-\Phi_v^\fl(X)= \\[6pt]
&=\na_vX+g(\na_v\t\,,X)\,\t-g(\t\,,X)\,\na_v\t
~:~\L\to\HL~,
\end{align*}
for any sections $v:\L\to\TO\L$ and $X:\L\to\HL$\,.

The usual \emph{Fermi derivative} is defined as a derivation
with respect to the detector's proper time, that is
$$\DO X\equiv\DO_{\t}X:\L\to\LL^{-1}\tn\HL~,\qquad
X:\L\to\HL~.$$

\begin{proposition}
For any $v:\L\to\TO\L$ and $X,Y:\L\to\HL$ one has
$$v.(X\cdot Y)=(\DO_vX)\cdot Y+X\cdot\DO_vY~.$$
\end{proposition}\proof
It follows from the fact that $\td\G$ is metric and $\Phi$ is anti-symmetric,
so that $\Phi^\fl$ is valued into the Lorentz group
(see~\Sec\ref{s:Group considerations} for more details about that).
We can directly verify our statement by observing that
the Lie derivative along $v$ of the scalar field $X\cdot Y$
is well-defined on $\L$ independently of extensions. We then have
\begin{align*}
v.(X\cdot Y)&=(\na_vX)\cdot Y+X\cdot\na_vY
=[\DO_vX+\Phi_v^\fl(X)]\cdot Y+X\cdot[\DO_vY+\Phi_v^\fl(Y)]=
\\[6pt]
&=(\DO_vX)\cdot Y+X\cdot\DO_vY+\Phi_v(X^\fl,Y^\fl)+\Phi_v(Y^\fl,X^\fl)
=(\DO_vX)\cdot Y+X\cdot\DO_vY~,
\end{align*}
since $\Phi_v$ is antisymmetric.
\qed

A section $X:\L\to\HL$ which is covariantly constant relatively to $\GF$ 
(namely $\DO X=0$\,, or $\DO_vX=0$ for all $v:\L\to\TO\L$)
is said to be \emph{Fermi-transported} along $\L$;
a Fermi-transported section is uniquely determined\footnote{
This follows from a well-known result about general connections,
since $\GF$ is a true connection on the restricted bundle $\HL\onto\L$
for fixed $\L$.}
by the value it takes at any point of $\L$.

A few points are worth stressing:
\begin{list}{$\bullet$}{}
\item
The scalar product of Fermi-transported vectors is constant along $\L$
(this follows at once from the above proposition).
\item
$\t$ itself is Fermi-transported;
if $f:\L\to\RR$ then $\DO(f\,\t)=(\t.f)\,\t$\,.
\item
If $X:\L\to\HL^\sbot$
(the subbundle of $\HL$ orthogonal to $\t$)
then also $\DO_vX:\L\to\HL^\sbot$,
coinciding with the orthogonal projection onto $\HL^\sbot$
of the ordinary covariant derivative $\na_vX$\,.
\end{list}

Thus $\GF$ preserves the splitting of $\HL$ into
the direct sum of its subbundles parallel and orthogonal to $\t$\,.
Moreover one also has Fermi-transported orthonormal frames $(\t_\l)$ of $\HL$
such that $\t_0\equiv\t$
(one only has to fix the frame at some point of $\L$ and `Fermi-transport' it).

In any orthonormal frame (not necessarily Fermi-transported)
one has the coordinate expressions
\begin{align*}
& \DO_v X=v^4\,\bigl(\de_4X^\l\,\t_\l-X^k\,\td\G\iIi4jk\,\t_j\bigr)\equiv
v.X^0\,\t_0+\bigl(v.X^j-v^4\,X^k\,\td\G\iIi4jk\bigr)\,\t_j~,
\\[6pt]
& \DO X=\Th_0^4\,\bigl(\de_4X^\l\,\t_\l-X^k\,\td\G\iIi4jk\,\t_j\bigr)~,
\end{align*}
which are independent of any extensions of $v$ and $X$\,.

\remark~
The definition of the Fermi derivative could be extended to the case
when $\L$ is spacelike, but cannot be immediately extended
to a derivative along a null 1-dimensional submanifold,\footnote{
Samuel and Nityananda~\cite{SN00}
have introduced a somewhat different transport law for polarization vectors
along non-geodesic null curves.}
since in the latter case there exists no normalized tangent vector ($\t$).
Moreover, the Fermi transport along an arbitrary timelike curve
cannot be seen as parallel transport
relatively to some connection on $\H\onto\M$.

However, a different kind of extension can be devised.
For this purpose, we first note that the section
$\Phi^\fl:\L\to\TS\L\tn\End(\HL)$
can be extended via the spacetime metric to a section
$$\Phi^\fl:\L\to\TOL^*\M\tn\End(\HL)~.$$
Namely we set
$$v\pint\Phi^\fl:=g(v,\t)\t\pint\Phi^\fl~,\quad v\in\TOL\M~.$$

Suppose that $\M$ is filled with congruence
of timelike 1-dimensional submanifolds,
with normalized tangent vector field $\t:\M\to\H$.
Then considering the above said extension for all said submanifolds
we obtain a section
$$\Phi^\fl:\M\to\TS\M\ten{\M}\End(\H)\cong\TS\M\ten{\M}\End(\TO\M)~.$$
Consider now the new spacetime connection $\td\G+\Phi^\fl$.
This has the property that the parallel transport along lines of the
chosen congruence coincides with Fermi transport there;
the same is \emph{not} true, however, for lines which do not belong
to the chosen congruence.
Also, note that the transport along spacelike lines
orthogonal to the lines of the congruence
coincides with ordinary parallel transport.

\subsection{Fermi transport of 2-spinors}
\label{s:Fermi transport of 2-spinors}

Let $\UL\onto\L$ be the restriction of the bundle $\U\onto\M$
to the base manifold $\L$\,.

Introducing an appropriate Fermi transport for spinors amounts essentially
to defining a modification $\CF$ of the connection\footnote{
For simplicity, we denote the restriction of $\Cs$ by the same symbol.}
$\Cs$ on $\UL\onto\L$,
in such a way that the induced connection $\CF\tn\bCF$
on $\HL\onto\L$ coincides with $\GF$\,.
The solution to the problem of determining $\CF$ is not unique,
as we'll see,
so one has got to describe the family of all solutions
(in the next section, the results obtained here will be extended to 4-spinors).

There is a natural procedure we can follow:
writing down an analogous of the relation between $\td\G$ and $\Cs$
(\Sec\ref{s:Two-spinor connections}).
We start from the two-spinor form of $\Phi^\fl$, namely
\begin{align*}
&\Phi^\fl=\Phi\iIi4\AAd\BBd\,\dx^4\tn\zeA\tn\bzeA\tn\zzB\tn\bzzB
:\M\to\TS\L\tn\End(\H)~,
\\[6pt]
\text{with}\quad
&\Phi\iIi4\AAd\BBd=\Phi\iIi4\l\m\,\t\iI\l\AAd\,\tt\Ii\m\BBd
=\oh\,\Phi\iIi4\l\m\,\s\iI\l\AAd\,\s\Ii\m\BBd~.
\end{align*}
By taking half the trace of $\Phi^\fl$
relatively to its conjugate 2-spinor indices
we get the section
$$\phi:\L\to\TS\L\ten{\L}\End(\UL)$$
which has the coordinate expression
$\phi=\phi\iIi4\sA\sB\,\dx^4\tn\zeA\tn\zzB$ with
$$\phi\iIi4\sA\sB=\oh\,\Phi\iIi4{\AAd}{\sB\cA}~.$$
A simple calculation, using the properties of the Pauli matrices,
gives then
$$\phi\iIi4\sA\sB=\oh\,\td\G\iI4{0j}\,\s\iIi j\sA\sB~.$$
Note that $$\Tr(\phi)=\phi\iIi4\sA\sA\,\dx^4=0$$
(in agreement with $\Phi\iIi4\l\l=\Phi\iIi4\AAd\AAd=0$).
Conversely, it's not difficult to show---by standard 2-spinor algebra---that
$$\Phi\iIi4\AAd\BBd=
\phi\iIi4\sA\sB\,\d\Ii\cA\cB+\d\Ii\sA\sB\,\bar\phi\iIi4\cA\cB~.$$

Next, we introduce the \emph{spinor Fermi connection} on $\UL\onto\L$\,,
$$\CF:=\Cs+\phi~,$$
which has the coordinate expression
\begin{align*}
(\CF)\iIi4\sA\sB=\Cs\iIi4\sA\sB+\phi\iIi4\sA\sB
&=(G_4+\iO\,Y_4)\,\d\Ii\sA\sB
+\oh\,\td\G\iIi4{\AAd}{\sB\cA}+\oh\,\Phi\iIi4{\AAd}{\sB\cA}
\\[6pt]
&=(G_4+\iO\,Y_4)\,\d\Ii\sA\sB+\oh\,(\GF)\iIi4{\AAd}{\sB\cA}~.
\end{align*}

If $v:\L\to\TO\L$ and $u:\L\to\UL$ are sections, then
$$ \na_v[\CF]u^\sA=\na_v[\Cs]u^\sA-v^4\,\phi\iIi4\sA\sB\,u^\sB
=v^4\,(\de_4u^\sA-\Cs\iIi4\sA\sB\,u^\sB
-\oh\,\td\G\iI4{0j}\,\s\iIi j\sA\sB\,u^\sB)~.$$

\begin{proposition}
The connection $\CF\tn\bCF$ induced by $\CF$ on $\HL\onto\L$
coincides with the the Fermi connection $\GF$\,.
Moreover,
any other linear connection $\CF'$ of $\UL\onto\L$ yielding $\GF$
differs from $\CF$ by a term of the type $\iO\,\a\tn\id$
with $\a:\L\to\TS\L$\,, namely
$$(\CF')\iIi4\sA\sB=(\CF)\iIi4\sA\sB+\iO\,\a_4\,\d\Ii\sA\sB~.$$
\end{proposition}
\proof
The coefficients of $\CF\tn\bCF$ are
\begin{align*}
(\CF\tn\bCF)\iIi4\AAd\BBd&=
\CF\iIi4\sA\sB\,\d\Ii\cA\cB+\d\Ii\sA\sB\,\bCF\iIi4\cA\cB
=(\Cs\iIi4\sA\sB+\phi\iIi4\sA\sB)\,\d\Ii\cA\cB
+\d\Ii\sA\sB\,(\bar\Cs\iIi4\cA\cB+\bar\phi\iIi4\cA\cB)=
\\[6pt]
&=(\Cs\iIi4\sA\sB\,\d\Ii\cA\cB+\d\Ii\sA\sB\,\bar\Cs\iIi4\cA\cB\,)
+(\phi\iIi4\sA\sB\,\d\Ii\cA\cB+\d\Ii\sA\sB\,\bar\phi\iIi4\cA\cB\,)=
\\[6pt]
&=\td\G\iIi4\AAd\BBd+\Phi\iIi4\AAd\BBd~.
\end{align*}

Now we observe that any other connection $\CF'$ of $\UL\onto\L$
can be written as $\CF+\Xi$\,,
with $\Xi:\L\to\TS\L\tn\End(\UL)$\,.
The condition that $\CF'$ yields $\GF$ can be written as
\begin{align*}
&\CF\iIi4\sA\sB\,\d\Ii\cA\cB+\d\Ii\sA\sB\,\bCF\iIi4\cA\cB=
(\CF\iIi4\sA\sB+\Xi\iIi4\sA\sB)\,\d\Ii\cA\cB
+\d\Ii\sA\sB\,(\bCF\iIi4\cA\cB+\bar\Xi\iIi4\cA\cB\,)~,
\\[6pt]
\Rightarrow\quad
&\Xi\iIi4\sA\sB\,\d\Ii\cA\cB+\d\Ii\sA\sB\,\bar\Xi\iIi4\cA\cB=0~.
\end{align*}
A short discussion
then shows that $\Xi\iIi4\sA\sB=\xi\,\d\Ii\sA\sB$ with $\xi:\L\to\iO\,\RR$\,.
\qed
\medbreak\noindent
{\bf Conclusion:}~we obtained a family of connections
of the restricted bundle $\UL\onto\L$\,.
Each element of the family yields the standard Fermi transport,
and is characterized by the arbitrary choice
of an imaginary function on $\L$\,.
$\CF$ is a distinguished element of the family,
so we see it as the natural generalization of Fermi transport to 2-spinors.

\subsection{Fermi transport of 4-spinors}
\label{s:Fermi transport of 4-spinors}

The coefficients of the connection induced by $\Cs$ on $\Ua\onto\M$
(namely the dual of $\bar\Cs$, see~\Sec\ref{s:Two-spinor connections}) are
$$\ost{{\bar\Cs}}\iI{a\cB}\cA=-\bar\Cs\iIi a\cA\cB~.$$
The couple $(\Cs,\ost{{\bar\Cs}})$ then constitutes the induced
(4-spinor) connection on the bundle $\U\oplus\Ua\equiv\W\onto\M$.
Its coefficients can be expressed~\cite{C00b} in the form
$$\Cs\iIi a\a\b=
\iO\,Y\!_a\,\d\Ii\a\b+\oq\,\td\G\iI a{\l\m}\,(\g_\l\,\g_\m)\Ii\a\b~,\quad
\a,\b=1,2,3,4~.$$
Its restriction to $\WL\onto\L$ can then be modified
in order to obtain a \emph{4-spinor Fermi connection},
that is the connection
$$(\CF\,,\, \obCF)=(\Cs{+}\phi\,,\,\ost{{\bar\Cs}}{-}\bar\phi^*)$$
obtained from $\CF$ by a similar procedure.
Namely, this new connection differs from $(\Cs,\ost{{\bar\Cs}})$
by the section
$$(\phi,-\bar\phi^*):\L\to\TS\L\tn\End(\WL)~,$$
where the transpose conjugate
$$\bar\phi^*:\L\to\TS\L\tn\Ua_{\!\!\Ll} \tn\Uc_{\!\!\Ll}
\equiv\TS\L\tn\End(\Ua_{\!\!\Ll})~,$$
has the coordinate expression
$$(\bar\phi^*)\iI{4\cB}\cA=\oh\,\bar\Phi\iIi4{\cA\sA}{\cB\sA}
=\oh\,\td\G\iI4{0j}\,\bar\s\iIi j\cA\cB~.$$

After some calculations we also find
$$(\phi,-\bar\phi^*)=\oq\,\hat\g(\Phi)
=\oq\,\td\G\iI4{0j}\dx^4\tn(\g_0\,\g_j-\g_j\,\g_0)~,$$
where $\hat\g:\wedge\H\to\End\W$ is the natural extension of the Dirac map.
For simplicity, let us indicate a connection on $\UL\onto\L$
and the induced connection on $\WL\onto\L$ by the same symbol;
then the induced 4-spinor Fermi connection of $\WL\onto\L$
can be written as
$$\CF=\Cs+(\phi,\,-\bar\phi^*)=\Cs+\oq\,\hat\g(\Phi)~.$$
Any other member $\CF{}'$ of the family of 4-spinor Fermi connections
are obtained from the above expression via the replacement
\hbox{$\phi\to\phi+\iO\,\a\tn\Id{\Uu}$}\,, namely
$$\CF{}'=\CF+\iO\,\a\tn\Id{\Ww}~.$$

\subsection{Group considerations}
\label{s:Group considerations}

A detailed study of the relations between 2-spinor groups,
4-spinor groups and the Lorentz group was exposed in~\cite{C07}.
In this section I'll just recall a few results which are relevant
in the present discussion.

In the algebraic setting of \Sec\ref{s:Two-spinor space}
consider the group
$$\SlG(\U):=\{K\in\Aut(\U):\det K=1\}~,$$
which preserves the two-spinor structure.
Its relations with the special orthochronous Lorentz group
and the orthochronous Spin group are described by the commutative diagram
$$
\begin{picture}(308,80)
\put(0,66){$\SlG(\U)$}
\put(36,69){\vector(1,0){42}}
\put(84,66){$\Lor_+^\up(\H)$}
\put(10,60){\vector(0,-1){47}}
\put(25,13){\vector(3,2){68}}
\put(0,0){$\Spin^\up(\W)$}
\put(158,35){\LARGE:}
\put(190,0){
\put(14,66){$K$}
\put(33,69){\vector(1,0){45}}
\put(84,66){$K\tn\bar K$}
\put(18,60){\vector(0,-1){47}}
\put(30,13){\vector(4,3){65}}
\put(0,0){$\bigl(K,(\bar K{}^\lin)^{-1}\bigr)$} }
\end{picture}
$$

One has the isomorphic Lie algebras $\Lie\Lor\equiv\Lie\Lor(\H)$\,,
$\Lie\Spin\equiv\Lie\Spin(\W)$ and
$$\Lie\SlG\equiv\Lie\SlG(\U)\cong\{\phi\in\End(\U):\Tr\phi=0\}~.$$
Furthermore $\Lie\Lor(\H)$ is isomorphic, as a vector space, to $\weu2\H$.
Thus one has the diagram of isomorphisms
$$
\begin{picture}(310,80)
\put(0,66){$\Lie\SlG(\U)$}
\put(65,69){\vector(1,0){25}} \put(65,69){\vector(-1,0){25}} 
\put(94,66){$\Lie\Lor(\H)$}
\put(15,35){\vector(0,-1){23}} \put(15,34){\vector(0,1){24}}
\put(75,3){\vector(1,0){21}} \put(76,3){\vector(-1,0){22}}
\put(0,0){$\Lie\Spin(\W)$}
\put(100,0){$\weu2\H$}
\put(110,35){\vector(0,-1){23}} \put(110,35){\vector(0,1){24}}
\put(170,35){\LARGE:}
\put(200,0){
\put(11,66){$\phi$}
\put(57,69){\vector(1,0){30}} \put(57,69){\vector(-1,0){30}} 
\put(97,66){$\Phi^\fl$}
\put(15,35){\vector(0,-1){23}} \put(15,34){\vector(0,1){24}}
\put(70,3){\vector(1,0){21}} \put(71,3){\vector(-1,0){22}}
\put(0,0){$(\phi,-\bar\phi{}^*)$}
\put(97,0){$\Phi$}
\put(101,35){\vector(0,-1){23}} \put(101,35){\vector(0,1){24}} }
\end{picture}
$$
where the relations among the above objects are as follows.

\medbreak\noindent
$\boldsymbol{a})$~$\Phi^\fl\in\Lie\Lor(\H)\subset\End(\H)=\H\tn\H^*$
is obtained from $\Phi\in\weu2\H\subset\H\tn\H$ through the isomorphism
$g^\fl:\H\to\H^*$ determined by the Lorentz metric.

\medbreak\noindent
$\boldsymbol{b})$~$\phi\in\Lie\SlG(\U)\subset\U\tn\Ul$
is one-half the trace of $\Phi^\fl$ relatively to the conjugate factors. Conversely, $\Phi^\fl=\phi\tn\Id{\Uc}+\Id{\U}\tn\bar\phi$\,.
Hence $\Tr\phi=0$\,.

\medbreak\noindent
$\boldsymbol{c})$~$\oq\,\hat\g(\Phi)=
(\phi,-\bar\phi{}^*)\in\End\U\oplus\End\Ua\subset\End\W$\,,
where $\hat\g:{\wedge}\H\to\End\W$ is the natural extension of the Dirac map
to the exterior algebra of $\H$.

\medbreak
Furthermore, it should be observed that
the biggest group which preserves the two-spinor structure is not $\SlG(\U)$
but rather the `complexified' group
$$\SlG^c(\U):=\{K\in\Aut(\U):|\det K|=1\}
=\bigl(\Ug(1)\times\SlG(\U)\bigr)/\ZZ_2~,$$
which leaves any symplectic form of $\U$ invariant up to a phase factor.
Its Lie algebra is
$$\Lie\SlG^c(\U)\cong\{A\in\End(\U):\Re\Tr A=0\}=\iO\,\RR\oplus\Lie\SlG(\U)~.$$
Accordingly, any $\th\in\Lie\SlG^c(\U)$ can be uniquely decomposed as
\medbreak\noindent
$\boldsymbol{d})$~
$\th=\oh\,(\Tr\th)\,\id+\bigl(\th-\oh\,(\Tr\th)\,\id\bigr)
\equiv \iO\,\a\,\id+\phi~,\quad \a\in\RR~,~~\phi\in\Lie\SlG(\U)~,$
\medbreak\noindent
with $\id\equiv\Id{\U}$\,, and one has
$$\th\tn\bar\id+\id\tn\bar\th=\phi\tn\bar\id+\id\tn\bar\phi~.$$
In other words, $\th\in\Lie\SlG^c(\U)$ determines an element
$\Phi^\fl\in\Lie\Lor(\H)$ via its traceless part $\phi$\,.

In the bundle setting of \Sec\ref{S:Two-spinor bundle and field theories}
the above spaces and groups become vector bundles and group bundles over $\M$.
Consider sections
\begin{align*}
& \phi:\M\to\TS\M\ten{\M}\Lie\SlG(\U)~,
\\[6pt]
& \th\equiv\iO\,\a\tn\Id{\U}+\phi:
\M\to\TS\M\ten{\M}\Lie\SlG^c(\U)~,
\\[6pt]
& \Phi^\fl:\M\to\TS\M\ten{\M}\Lie\Lor(\H)~,
\\[6pt]
& \oq\,\hat\g(\Phi)=(\phi,-\bar\phi{}^\lin):\M\to\TS\M\ten{\M}\Lie\Spin(\W)~,
\end{align*}
fulfilling the same mutual relations $\boldsymbol{a}$, $\boldsymbol{b}$,
$\boldsymbol{c}$ and $\boldsymbol{d}$ as the previously considered
algebraic objects with the same names
($\a:\M\to\TS\M$ is now a real 1-form).

Such Lie-algebra-bundle valued 1-forms can be seen as
\emph{differences between linear connections
preserving the respective vector bundle structures}
(while the curvature tensors are \emph{2-forms} valued in the same
Lie-algebra-bundles).
More precisely,
it's not difficult to prove:
\begin{proposition}\label{p:connectiondifferences}
Let $\Cs$ and $\Cs'$ be 2-spinor connections,
and $\td\G$, $\td\G'$ the respectively induced connections of $\H\to\M$.
Then
$$\th\equiv\iO\,\a\tn\id+\phi:=\Cs{-}\Cs':\M\to\TS\M\ten{\M}\Lie\SlG^c(\U)$$
and
$$\Phi^\fl:=\td\G{-}\td\G':\M\to\TS\M\ten{\M}\Lie\Lor(\H)$$
fulfil the above relations
$\boldsymbol{a}$,\,$\boldsymbol{b}$,\,$\boldsymbol{c}$,\,$\boldsymbol{d}$.
In particular, $\Phi^\fl$ only depends on the traceless part $\phi$ of $\th$\,.
\end{proposition}

\section{An application: free QED states}
\label{S:An application: free QED states}

Though a kind of `covariance' can be achieved in flat spacetime,
current quantum theory remains essentially observer-dependent.
This feature is most evident when one tries to formulate QFT
in curved spacetime,
where one is unable to define a distinguished,
observer-independent set of free states
(\eg\ see Birrel and Davies~\cite{BD}).

In a previous paper~\cite{C05} I studied a quantum formalism,
in momentum space,
carried by a pointlike observer:
connections on underlying `classical' bundles
determine `quantum connections' on `distributional bundles',
namely bundles over spacetime whose fibres are distributional spaces,
and restriction to a given observer worldline does the job.\footnote{
At least locally the chosen worldline determines, vie exponentiation,
a splitting space$\,{+}\,$time enabling a position space representation.}
I also hinted at the possibility that free electron states,
for the observer's formalism,
be described in terms of a Fermi transport of spinors,
rather than by ordinary covariant transport.
This seems natural in view of the standard interpretation
of the usual Fermi transport of vectors
in relation to small gyroscopes carried by the observer.

For $m\in\LL^{-1}$ let $\Pm\subset\TS\M$
be the subbundle over $\M$ of all future-pointing $p\in\TS\M$
such that $g^\#(p,p)=m^2$\,.
Then $\Pm$ is the classical \emph{momentum bundle} for a particle of mass $m$
(the limit case $m=0$ can also be considered).
Consider the 2-fibred bundle
$$\bigl[(\weu3\TS\Pm)^+\bigr]{}^{1/2}\equiv\VV^{-1/2}\Pm\onto\Pm\onto\M$$
whose upper fibres are the spaces of \emph{half-densities}
of the momentum spaces.
If
$\V\onto\Pm$ is a complex vector bundle
(whose fibres represent the \emph{internal degrees of freedom} of the particle)
then for each $x\in\M$ one has
the vector spaces $\VC^1_{\!x}$ of all \emph{generalized sections}
(in a distributional sense)
$$(\Pm)_x \gto (\VV^{-1/2}\Pm\ten{\Pm}\V)_x~,$$
which can be assembled~\cite{C04a} into a smooth bundle $\VC^1\onto\M$
(smoothness being defined in a certain, appropriate way).
A \emph{Fock bundle} can be constructed as
$\VC:=\bigoplus_{i=0}^\infty\VC^i$\,,
where $\VC^i$ is defined to be either
$\weu{i}\VC^1$ or $\vee^i\VC^1$
(respectively, antisymmetrized and symmetrized tensor products
for fermions and bosons).
Thus one particle states are represented as $\V$-valued
\emph{generalized half densities}.

The spacetime connection determines a connection $\G_{\!\!m}$ on $\Pm\onto\M$\,;
moreover, in the usual physical situations one has a connection
$\V\to\Pm\to\M$ which is linear projectable over $\G_{\!\!m}$\,.
These determine a connection on $\VC^1\onto\M$ and hence on $\VC\onto\M$.

For each $p\in(\Pm)_x$\,, $x\in\M$,
let $\d_p$ be the Dirac density with support $\{p\}$ in $(\Pm)_x$\,,
$\om_m$ the \emph{Leray density} of $(\Pm)_x\subset\TO_{\!x}^*\M$
and $\bigl(\bb_\a(p)\bigr)$ a basis of $\V_{\!\!p}$\,.
Let moreover $l\in\LL$ be a `length unit'.
Then the set $\bigl(\Bsf_{p\a}\bigr)$\,, with
$$\Bsf_{p\sA}:=\frac1{\sqrt{2\,l^3\,p_0}}\,
\d[p]\tn\om_m^{-1/2}\tn\bb_\a~,$$
constitutes a \emph{generalized frame} of $\VC^1$.
The above said connection on $\VC\onto\M$
yields parallel transport of such frames along curves in $\M$\,;
in particular, 4-momentum $p$ is parallely transported.

Consider now the bundle $\W\onto\M$ of Dirac spinors.
For each $p\in\Pm$ one has a splitting\footnote{
The restrictions of the Hermitian metric $\kO$
(\Sec\ref{s:From 2-spinors to Dirac spinors})
to these two subspaces turn out to have the signatures
\hbox{$(+,+)$} and \hbox{$(-,-)$}\,, respectively.}
$$\W=\W_{\!\!p}^+\dir{\M}\W_{\!\!p}^-~,\quad
\W_{\!\!p}^\pm:=\Ker(\g[p]\mp m)~.$$
Thus for each $m\in\LL^{-1}$ one has
the 2-fibred bundles $\W_{\!\!m}^\pm\to\Pm\to\M$ defined by
$$\W_{\!\!m}^\pm:=\bigsqcup_{p\in\Pm}\W_{\!\!p}^\pm\subset\Pm\cart{\M}\W~.$$
$\W_{\!\!m}^+$ and $\overline\W{}_{\!\!m}^-$ are then the \emph{electron bundle}
and the \emph{positron bundle}, respectively.

All the above constructions can be restricted to timelike one-dimensional
base manifold $\L\subset\M$. 
In order to introduce appropriate generalized frames
for free electron and positron states along $\L$
one needs, for each $p\in(\Pm)_\Ll$\, a frame
$$\bigl(\uu\!_\sA(p)\,,\,\vv\!_\sA(p)\bigr)~,\quad{\scriptstyle A}=1,2~$$
of $\W_{\!\!p}$ which is adapted to the splitting
$\W_{\!\!p}=\W_{\!\!p}^+\oplus\W_{\!\!p}^-$\,.
A consistent choice can be made by the following procedure.

Fix any point $x_0\in\L$. Let $\t_{x_0}\in\LL^{-1}\tn\TO_{\!x_0}\L$
be the unit future-pointing scaled vector field.
Choose any 2-spinor basis $\bigl(\zeA\bigr)$
such that the timelike element $\t_0$
of the induced Pauli basis coincides with $\t_{x_0}$\,.
Then the corresponding Dirac basis
(\Sec\ref{s:From 2-spinors to Dirac spinors}),
constituted by the elements
$$\uu_1:=\osq\,(\ze_1\,,\bzz^1)\;,~
\uu_2:=\osq\,(\ze_2\,,\bzz^2)\;,~
\vv_1:=\osq\,(\ze_1\,,-\bzz^1)\;,~
\vv_2:=\osq\,(\ze_2\,,-\bzz^2)\;,$$
is adapted to the splitting determined by $p\equiv m\,g^\fl(\t_{x_0})$\,.
Next we Fermi-transport this basis along $\L$.
$\t$ itself is Fermi-transported,
and so is the 1-form $\t^\fl:=g^\fl(\t)$ corresponding to $\t$
via the spacetime metric.
Thus we get a Dirac frame
$\bigl(\uu\!_\sA(m\,\t^\fl)\,,\,\vv\!_\sA(m\,\t^\fl)\bigr)$
which is adapted to splitting determined by $p\equiv m\,\t^\fl$.

Now we have to extend this to a Dirac frame of $\WL$ for all $p\in(\Pm)_\Ll$\,.
This can be done, at each $x\in\L$,
essentially by the usual procedure of the flat inertial case.
Namely, if $p\in(\Pm)_x$ is now an arbitrary 4-momentum at $x$ then
take the unique boost $\Lambda$ such that $\Lambda(\t_x)=g^\#(p)/m$\,;
up to sign there is a unique transformation $K\in\Spin(\W)$
which projects over $\Lambda$,
and an overall sign can be fixed by continuity.\footnote{
This is related to the structure of boosts.
See~\cite{C07} for a detailed account of the relations
among spinor groups and the Lorentz group in terms of 2-spinors.}
This $K$ transforms the Dirac frame
$\bigl(\uu\!_\sA(m\,\t^\fl)\,,\,\vv\!_\sA(m\,\t^\fl)\bigr)$
into the new Dirac frame
$\bigl(\uu\!_\sA(p)\,,\,\vv\!_\sA(p)\bigr)$\,.

The introduction of free photon states has subtilities of a different nature,
while their transport along the observer's world line
is performed via ordinary Fermi transport.
Then one modifies the induced \emph{free-particle connection}
on the Fock bundle of QED by an \emph{interaction}
(not deduced from any underlying classical structure)
which yields the full picture of electrodynamics
(see~\cite{C05} for details).


\end{document}